\documentclass[11pt,a4paper]{amsart}

\usepackage[utf8]{inputenc}
\usepackage{amsthm,amsfonts,amsmath,amssymb,graphicx, color}
\usepackage[colorlinks=true,citecolor=black,linkcolor=black,urlcolor=blue]{hyperref}
\usepackage[numbers,sort&compress]{natbib}

\usepackage{algorithmic}
\usepackage{algorithm}

\allowdisplaybreaks

\renewcommand{\leq}{\leqslant}

\theoremstyle{plain}

\DeclareMathOperator{\R}{\mathbb{R}}

\newcommand{\bc}{\begin{center}}
\newcommand{\ec}{\end{center}}

\begin{document}
%%%%%%%%%%%%%%%%

\title[OOVD and the $1$-Steiner tree problem]{Overlaid oriented Voronoi diagrams and the $1$-Steiner tree problem}
\author{Michael S. Payne, Charl Ras, Marcus Volz}

\maketitle

\begin{abstract}
Overlaid oriented Voronoi diagrams (OOVDs) are known to provide useful data for the construction of optimal Euclidean $1$-Steiner trees.
The theoretical time complexity of construction methods exploiting the OOVD is $O(n^2)$, but a computational study has never been performed, and robust constructions for OOVDs have not previously been implemented.

%Although there exist theoretical bounds for the complexity of constructing OOVDs, and for the reduction in complexity when using OOVD-based $1$-Steiner tree construction methods versus naive methods,

In this paper, we outline a numerically stable implementation for constructing OOVDs using tools from the Computational Geometry Algorithms Library (CGAL), and test its performance on random point sets.
%
%We perform a computational study to show that our algorithm is fast and robust.
%
%Our experiments reveal that the average structural complexity of an OOVD is approximately $O(n)$, where $n$ is the number of input terminals -- which is significantly less than the proven $O(n^2)$ worst case. 
%
We then study the effect that the OOVD data has in reducing the complexity of $1$-Steiner tree construction when compared to a naive approach.
The number of iterations of the main loop of the 1-Steiner algorithm is directly determined by the number of faces in the OOVD, and this appears to be linear for the random inputs we tested.
We also discuss methods for processing the OOVD data that lead to a reduction in %of $1$-Steiner tree 
construction time by roughly a factor of 12.
%
%Finally, we provide a number of new statistical observations about the structure of optimal Euclidean $1$-Steiner trees.
\end{abstract}

\section{Introduction}

The Euclidean Steiner tree problem is to construct the network with shortest total edge length that connects a set of input terminals in the plane. The leading algorithm for this problem, GeoSteiner, has at its core a pruning technique that uses the fact that any Euclidean minimum spanning tree (MST) on a set of points $P$ is a subgraph of the \emph{relative neighbourhood graph} of $P$~\cite{Toussaint80, JuhlWZ18, WarmeWZ00}. % (WHAT REF FOR GEOSTEINER?).
The relative neighbourhood graph is a subgraph of the Delaunay triangulation of $P$, and is therefore related to its dual, the Voronoi diagram of $P$. Optimal Euclidean Steiner trees contain new junctions, or \emph{Steiner points}, which all have degree 3.
 
The Euclidean $k$-Steiner problem on $P$ is to construct the network with shortest total edge length that connects the points of $P$ and uses at most $k$ Steiner points.
This problem requires a different approach to the original Steiner tree problem for the simple reason that Steiner points of degree $4$ can occur in an optimal solution.
The success of the GeoSteiner algorithm in efficiently solving large Steiner tree problems is crucially tied (via the Melzak algorithm) to the assumption that all Steiner points are of degree $3$.

The Steiner tree problem is NP-hard, while it is fairly easy to show that the $k$-Steiner tree problem is solvable in polynomial time for constant $k$. %~\cite{BrazilRST15}
That said, it takes significant ingenuity to keep the degree of the polynomial reasonably low.
The first result in this direction was a quadratic time algorithm for the $k=1$ case described by Georgakopoulos and Papadimitriou in  1987~\cite{GeorgeP87}. In that paper they demonstrated the utility of another kind of Voronoi diagram, the overlaid oriented Voronoi diagram.

\begin{figure}
    \centering
    \includegraphics[width=\textwidth]{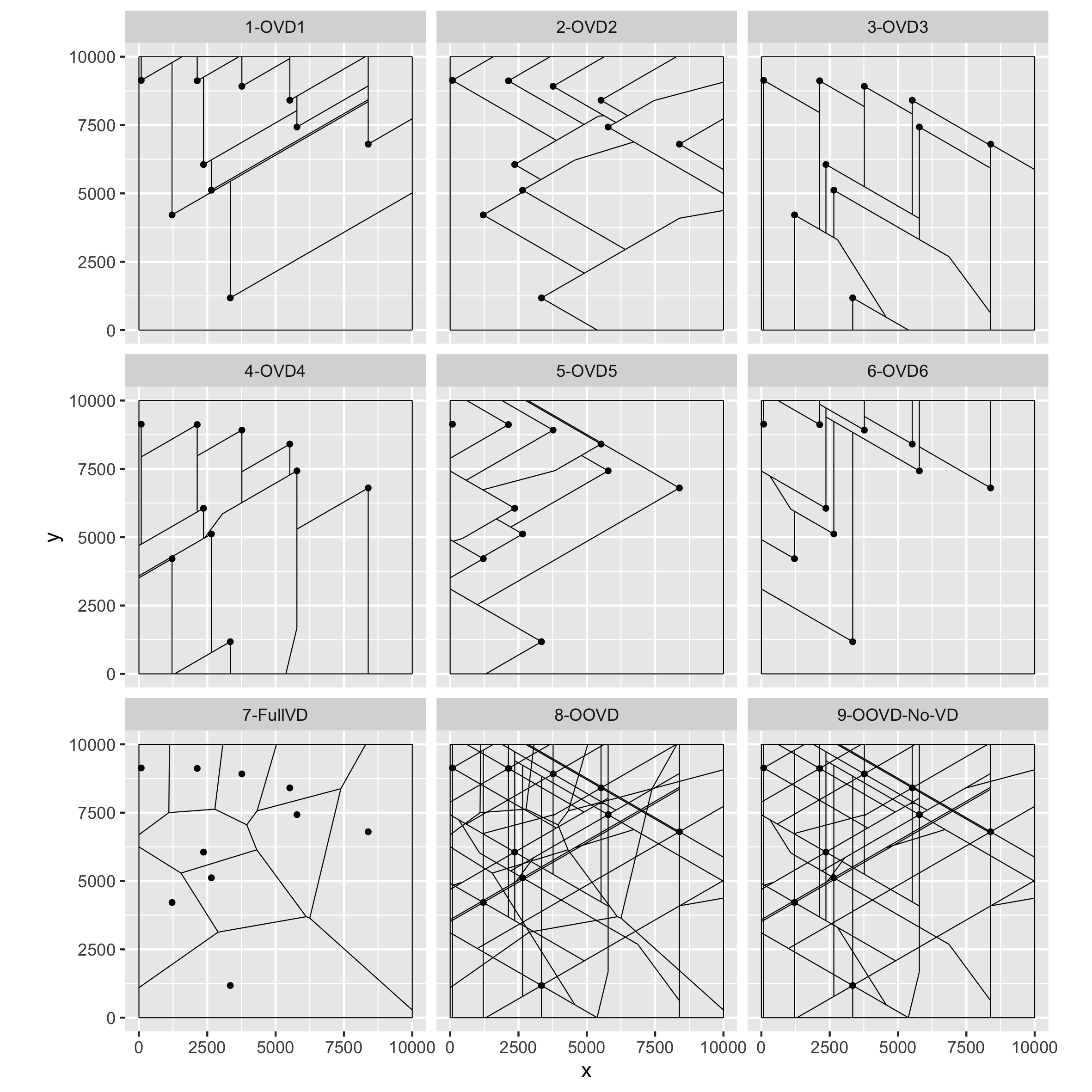}
    \caption{An example of an OOVD for 10 terminals.}
    \label{fig:oovd}
\end{figure}

Recall that the classical Voronoi diagram of a set $P$ essentially records, for each point in the plane, the closest point in $P$.
An \emph{oriented Voronoi diagram} (OVD) for $P$ and a given cone $C$ records, for each point $x$ in the plane, the closest point of $P$ within the translate of $C$ based at $x$. We think of the cone $C$ informally as a `range of directios', e.g. $[0, \pi / 3]$.
The \emph{overlaid oriented Voronoi diagram} (OOVD) is the overlay of six OVDs corresponding to six equiangular cones that partition the plane.
Thus it records, for each point $x$, the closest point of $P$ in each of six different ranges of directions covering the whole plane. Examples of OVDs and an OOVD can be seen in Figure~\ref{fig:oovd}.

Monma and Suri~\cite{MonmaS92} independently discovered the power of the OVD when studying so-called `transitions' in minimum spanning trees. The question they asked is: how is a minimum spanning tree topology affected by perturbations of the the input terminals? The OVD structure allowed them to bound the number of minimum spanning tree topologies that can result when one of the terminals is perturbed. The problem of efficiently predicting topological changes of minimum spanning trees is important for sensitivity analysis, for instance in cases where the input terminal locations are not known with certainty. Models for transitions in MSTs also capture the topological changes in dynamic networks where, for example, input terminals are mobile. % [cite WHAT PAPER?].

A number of papers have since provided further theoretical results about the algorithmic construction of OVDs and their application to network design. In~\cite{ChangHT90} an $O(n\log n)$ sweep algorithm is presented for constructing the OVD. Note however, this does not provide an overall reduction in complexity of OOVD construction, since the overlay process still requires $O(n^2)$ time. In~\cite{BrazilRST15} OOVDs were applied to the more general problem of optimal $k$-Steiner tree construction in various planar norms and for various cost functions on the edges, yielding an algorithm of complexity $O(f(k)n^{2k})$, for some function $f$. The OOVD structure has also been applied to the design of survivable networks~\cite{Ras15}.

Despite the importance of OOVDs, we are not aware of any practical implementations or experimental studies of OOVD construction.
In Section~\ref{ovd} we describe the mathematical construction of the OVD that we use, discuss the CGAL code that implements it and creates the OOVD, and report on the performance of the code on random input sets. We find that the code is robust and appears to run in much better than the worst case quadratic time complexity on these inputs. The complexity of the OOVD itself seems essentially linear.

The $O(n^2)$ time algorithm for the 1-Steiner problem presented in~\cite{GeorgeP87} is theoretical and hasn't been refined or implemented practically.
In Section~\ref{Steiner} we discuss the 1-Steiner problem and the way the OOVD data helps to solve it. We use new pruning methods to reduce the OOVD data that result in roughly a factor of 12 improvement in running time over a naive implementation, and also make some statistical observations about the optimal 1-Steiner trees found during our experiment. The efficiency of the algorithm is closely tied to the complexity of the OOVD, and thus also appears to be much better than the worst case analysis suggests.

\section{The overlaid oriented Voronoi diagram}\label{ovd}%{Computing the OOVD}
% This section will:
% - summarise the maths behind OVD
% - describe the CGAL code at a high level
% - discuss issues and design choices
% It shouldn't go into the combinatorics (this isn't really relevant for k=1)
% Maybe leave complexity discussion until later, and discuss the whole algorithm at once.

% The section headings need reworking -- do it when content is more organised.

% Intro

%\subsection{Oriented Voronoi diagrams}

Given a set of input points, the oriented Voronoi diagram is a partition of the plane into cells that records, for each point in the plane, the closest input point in a fixed range of directions. 
To define the OVD formally, let $C$ be a fixed cone (based at the origin) 
and consider a set $P$ of input points. Then the cell of the OVD corresponding to the point $p\in P$ is $$ \{x\in \R^2 : x-p\in C \ \& \ (\forall q \in P: x-q \in C \Rightarrow |x-p| \leq |x-q|)  \}$$
%I know this is ugly. A multipart definition would be easier to understand.
Alternatively, we can define the cell in terms of translated cones $C_q$ based at $q$, and halfplanes $H_{qp}$ of points closer to $q$ than $p$.
Then the cell corresponding to $p$ is $$C_p \setminus \bigcup_{q\in P\setminus \{p\}} C_q \cap H_{qp}.$$
%
%As we will see, this second description is most relevant to the method we used to construct the OVDs.
Note that unlike Voronoi diagrams, the cells of an OVD are not always convex and do not cover the entire plane.

%KEEP ALL STEINER TREE DISCUSSION FOR THE STEINER TREE SECTION
%M: I think this paragraph helps to motivate the construction.
In the case of the $k$-Steiner problem, we use six OVDs corresponding to six cones of angle $\pi /3$ that partition the plane. 
The overlaid oriented Voronoi diagram is the common refinement of these six OVDs. Therefore each cell of the OOVD is the intersection of a cell from each of the six OVDs. 
%As such, it corresponds to up to six of the input points, one in each of the six cones. 
Thus it records the closest input point in each of the six ranges of directions determined by the cones.

%%%%%%%%%%%%%%
\subsection{Constructing the oriented Voronoi diagrams}

We use a very general method for calculating Voronoi type diagrams that uses the upper envelope of a set of surfaces in 3 space~\cite{EdelsGS89}.
Given a set of surfaces in $\R^3$, considered as graphs of functions of $x$ and $y$, the \emph{upper envelope} is a partition of the $xy$-plane according to which surface is highest above each point $(x,y)$. In other words, it is a planar map with each region corresponding to a subset of the domain for which a particular function has the highest value.

It is well known that the classical Voronoi diagram can be constructed as the upper envelope of a set of planes tangent to the unit paraboloid, $f(x,y) = x^2 + y^2$ \cite{EdelsS86, EdelsGS89}.
Consider a point $p=(p_x, p_y) \in \R^2$.
The tangent plane to $f$ at $p$ has equation $h_p(x,y) = 2p_x x + 2p_y y - p_x^2 - p_y^2 $.
Observe that at another point $q=(q_x,q_y)$, the vertical distance between $f(q)$ and $h_p(q)$ is
$$ f(q) - h_p(q) = q_x^2 + q_y^2 - 2p_xq_x - 2p_yq_y + p_x^2 + q_x^2 = (q_x - p_x)^2 + (q_y - p_y)^2.$$
That is, it is the square of the distance from $p$ to $q$.
Therefore, given a set of points $P= \{p_1, \ldots, p_i \}$, if one considers the set of tangent planes $h_{p_i}$, then for any point $q$ the plane that has the highest value at $q$ corresponds to the closest point in $P$ to $q$.
This means the classical Voronoi diagram of $P$ is precisely the upper envelope diagram for the planes $h_{p_i}$.

The OVD can be constructed in the same way by using a cone in each of the tangent planes $h_{p_i}$ with apex at $f(p_i)$.
In practice we use a triangle which is this cone truncated at one of the lines bounding the problem domain.
After constructing these triangles we calculate the upper envelope to create the OVD.
Once the OVDs are constructed, overlaying them is mathematically straightforward.

For reasons that we discuss in Section~\ref{Steiner}, we also construct the classical Voronoi diagram.
We use the same triangle envelope algorithm, but instead of using tangent planes, we create large tangent triangles that cover the entire domain. We will overlay the OOVD and the classical Voronoi diagram to create what we call the \emph{refined OOVD}.

Figure \ref{fig:oovd} shows an example of the six OVDs and the OOVD for a set of 10 input points. It also includes the classical Voronoi diagram, the refined OOVD, and the standard OOVD.

%The methods used to record and keep track of the relevant input terminals for each face of the OOVDs are discussed in the next section.

%%%%%%
\subsection{Overview of the CGAL code}

We decided to use the Computational Geometry Algorithms Library (CGAL) for OOVD construction because it has robust and fast implementations of the main geometric procedures we require, namely an upper envelope function for sets of triangles in $\R^3$ and an overlay function that returns the common refinement of two planar maps.
These functions are discussed in great detail in~\cite{FogelHW12}.

The following is a brief overview of our CGAL code for OOVD creation. We defer detailed discussion of the code until we are able to release it publicly.
From the input points, our CGAL program first creates the triangles required for the upper envelopes. To these triangles we attach data that records the associated input point. 
Next, the seven envelope diagrams are created. The data from the triangles is passed to the regions of the diagrams, associating each region with the relevant input point.
After trimming the diagrams down to the bounding box of the problem domain, we begin overlaying them. The overlay function overlays two diagrams at a time, so we need 6 overlay operations to combine them all.
During each overlay operation, the data from the faces of the input diagrams is combined to create the data for each face of the output diagram. 

The data is only valid on full dimensional faces, not edges or vertices.
The data from the faces of the OOVD, rather than the diagram itself, is what is used in the $k$-Steiner algorithm.

\subsection{Numerical accuracy and coordinate number type}

The validity of the data produced by our program depends on the faces of the OOVD being calculated precisely. 
Even with rational input points, the OVDs contain a lot of edges with irrational slopes, due to the $\pi / 3$ angles of the cones. Representing the diagrams with rational coordinates (including floating point) resulted in rounding errors that caused extra faces to appear -- essentially long thin faces that should have been edges. Since the number of faces directly affects the complexity of the $k$-Steiner algorithm, this was a significant issue.

One solution to the problem is to use coordinates in the quadratic extension field $\mathbb{Q}[\sqrt{3}]$.
Geometrically, since computers inevitably use bounded precision to represent reals, one can think of this as using coordinates in a triangular rather than square grid.
It can be shown that, for rational input points, all intersections of lines in the OVDs and their overlays can be represented exactly in these coordinates.
Fortunately CGAL comes equipped with the ability to use coordinates in a quadratic extension field. 
Our experiments confirmed that this approach eliminated the errors observed using rational coordinates.

\subsection{Theoretical complexity of computing the OOVD}\label{subOVDcomp}

Constructing the OOVD is part of the preprocessing, and is independent of $k$ for the $k$-Steiner problem. The $1$-Steiner problem can be solved in $O(n^2)$, while the CGAL algorithms that we use have theoretical running times slightly more than $O(n^2)$. However this turns out to be a very minor issue, as we will see.
For $k>1$ the theoretical bounds on the complexity of the algorithms we use are much lower than the $O(n^{2k})$ complexity of the $k$-Steiner algorithm.

The geometric complexity (number of faces, edges and vertices) of an OVD is $\Theta(n)$~\cite{GeorgeP87}, however the complexity of \emph{constructing} the OVD is $\Theta (n \log n)$, using a sweepline type algorithm~\cite{ChangHT90}.
In general, the upper envelope based method we use for computing the OVD has time complexity $\Theta(n^2 \alpha (n))$, where $\alpha (n)$ is the inverse Ackermann function~\cite{EdelsGS89}.
This bound is the same as the worst case geometric complexity of the upper envelope of a set of triangles, so there can't be a faster construction for the upper envelope of triangles in general. The algorithm used in CGAL may have even worse complexity of $O(n^{2+\epsilon})$ because it accepts collections of more general surfaces\footnote{Actually semi-algebraic surfaces of constant descriptive complexity.}~\cite{AgarwalSS96}.
However, as the complexity of an OVD is much less than this bound, we expect that the true running time of the upper envelope algorithm for the special case of OVD construction is significantly better than $O(n^2 \alpha (n))$.
Of course it can't be better than the $\Omega(n \log n)$ lower bound in~\cite{ChangHT90} which is due to a reduction from the sorting problem.
%Second, there is good reason to suspect that CGAL algorithms (which are very general) are actually much better than their worst case complexity suggests when run on the particular inputs used for OOVD construction.

Turning to the task of overlaying the six OVDs, the geometric complexity of the OOVD is $\Theta(n^2)$ in the worst case, and Georgakopoulos and Papadimitriou describe a straightforward algorithm to construct it in time $O(n^2)$~\cite{GeorgeP87}. They use the observation that we can reduce the question to the problem of overlaying triangulations, since the OVDs can be triangulated with $O(n)$ triangles.
Even without this simplification, it is known that the complexity of the overlay of a constant number of upper envelope diagrams of $n$ surfaces in $\R^3$ is $O(n^{2+\epsilon})$~\cite{AgarwalSS96}.
Thus the CGAL overlay algorithm may again have slightly worse than quadratic complexity in theory.

%Given two envelope diagrams with $m$ faces each, it is possible to construct their overlay in time $O(m^{2+\epsilon})$ for any $\epsilon >0$ [Agarwal Schwarzkopf Sharir 96].

\subsection{Experimental setup and results}\label{subOVDExp}

In practice the CGAL based implementation of OOVD construction worked very effectively.

\begin{figure}
    \centering
    \includegraphics[width=\textwidth]{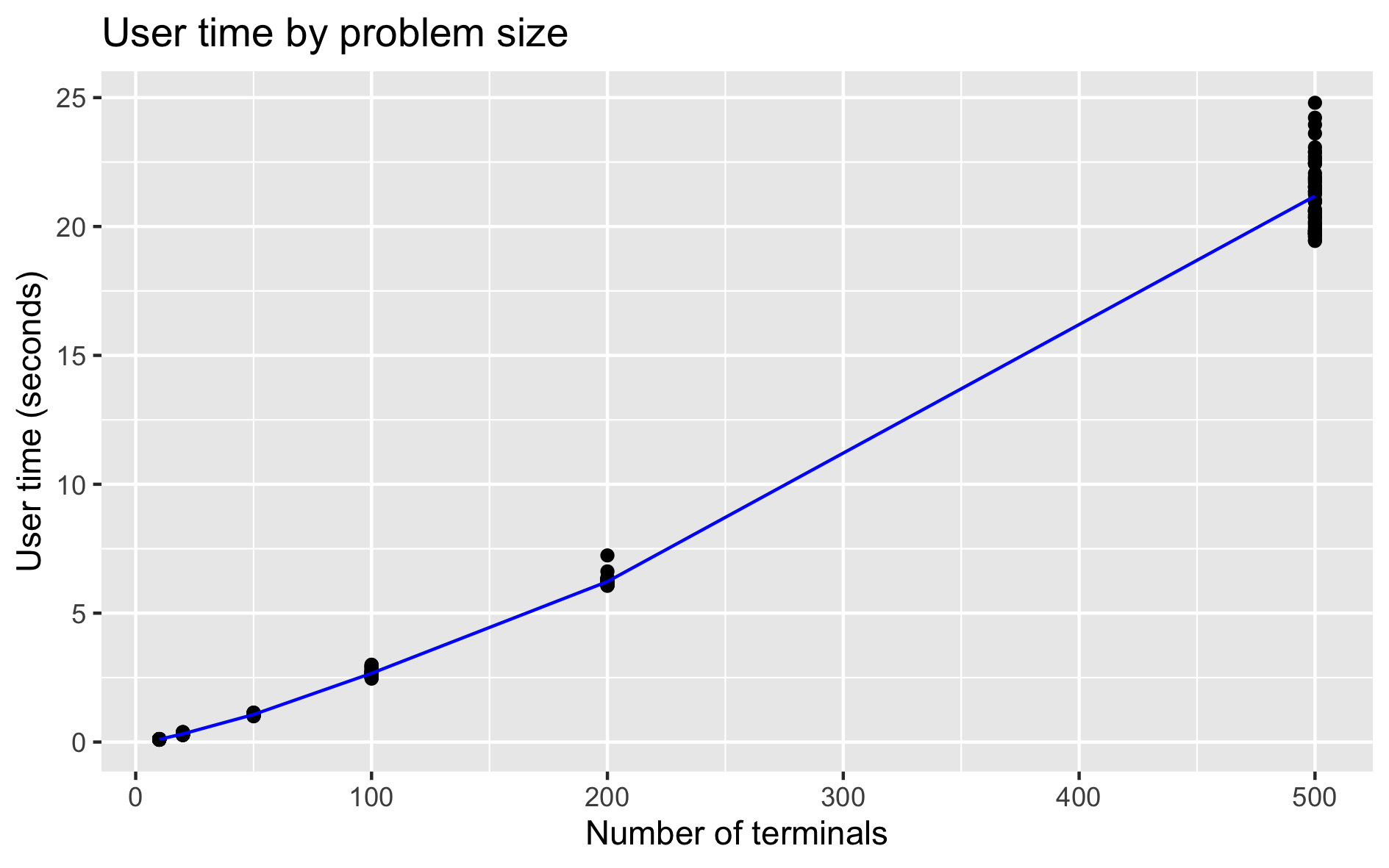}
    \caption{User time by problem size (averages of 50 seeds).}
    \label{fig:user-time}
\end{figure}

The OOVD code was tested on uniformly randomly generated point sets in a 10,000 by 10,000 integer grid. The main experiment involved computing OOVDs for sets of size 10, 20, 50, 100, 200 and 500, with 50 instances of each size. In each case we computed the six OVDs, the classical Voronoi diagram, the OOVD, the refined OOVD, and the output was just the data from the faces of the refined OOVD.
Figure~\ref{fig:user-time} shows the user time for all these instances and a line connecting the averages. All OOVD computations were performed on a system with an Intel Core i7-8650U CPU at 1.9 GHz (4.2GHz turbo). The system had 16GB of RAM, although very little was used.

The shape of the curve in Figure~\ref{fig:user-time} could suggest that on this kind of example the algorithm runs faster than the worst case $\Omega(n^{2+\epsilon})$ time discussed in Section~\ref{subOVDcomp}. However there is also a wide variation in time taken, making it hard to draw strong conclusions. For example, the 500 point examples took between 19 and 25 seconds to compute.

\begin{figure}
    \centering
    \includegraphics[width=\textwidth]{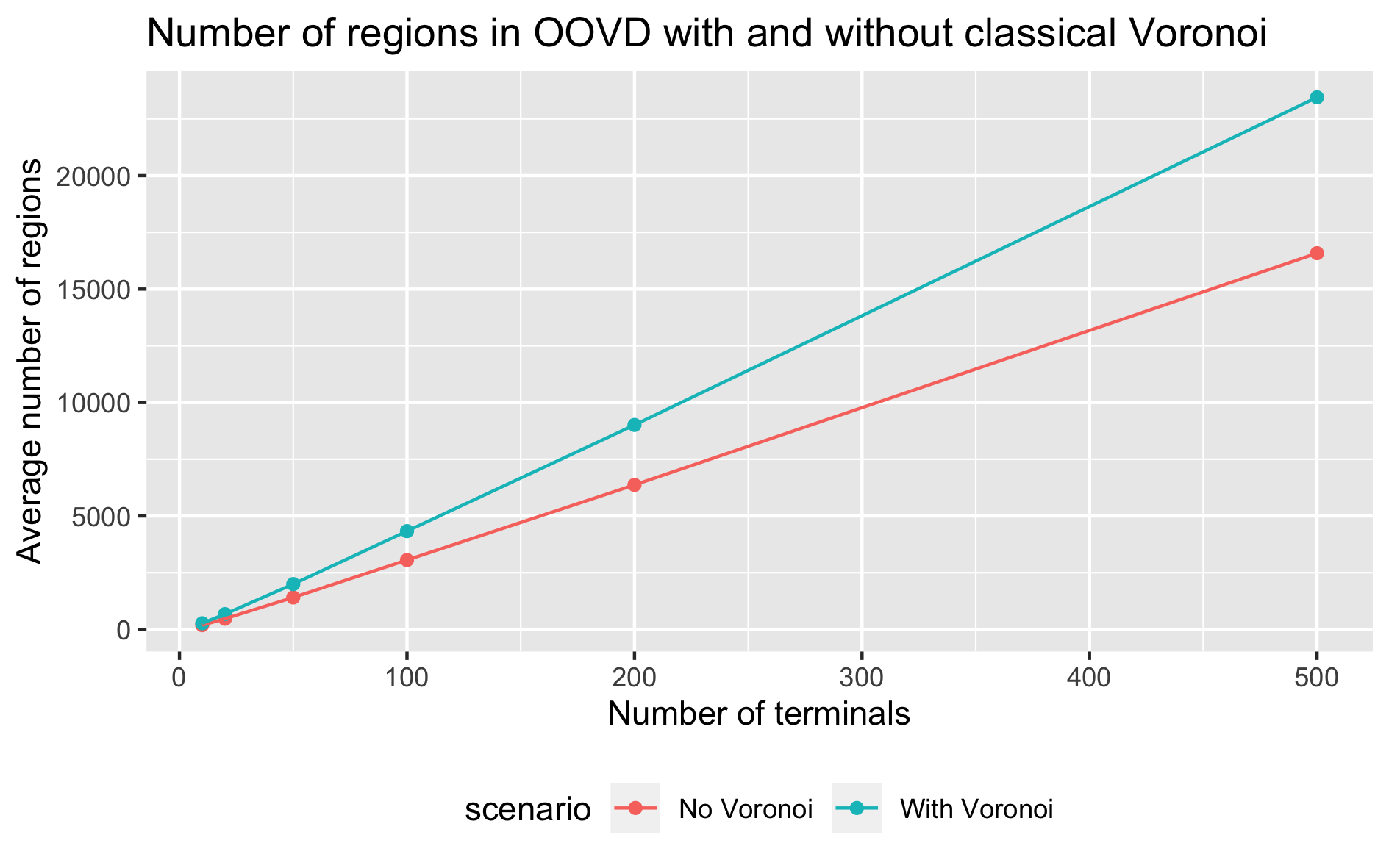}
    \caption{}
    \label{fig:regions}
\end{figure}

Overlaying the classical Voronoi diagram with the OOVD obviously increases the number of faces, and thus the amount of data to be used later. To illustrate the effect of this, we compared the average number of faces in the OOVD with and without the Voronoi overlay\footnote{Note however that number of faces in the OOVD alone was only calculated on a smaller set of examples.}. The resulting plot in Figure~\ref{fig:regions} shows a strikingly linear relationship between the number of input terminals and the number of regions in the OOVD. For example the number of cells in the refined OOVD is a little under 50 times the number of terminals. This is perhaps not surprising given the points were selected uniformly at random within the square domain. One could expect that the diagram looks locally similar everywhere regardless of the number of points (except near the domain boundary), so that the typical number of cells near a terminal is fairly constant. Such local behaviour would lead to the linear relationship that was observed.

It is feasible to calculate much larger instances on the same hardware. For example, one run on an input set of 10,000 points was completed in just over 17 minutes. This produced a refined OOVD with 483,000 regions, supporting the apparently linear relation in Figure~\ref{fig:regions}. The computation used almost 3GB of RAM at one point.

In summary, our results thus far suggest that, for the kind of input we used, the number of regions in the OOVD is linear and the time taken for construction is not much worse than linear.
Clearly further experiments could be run on larger examples to explore the performance of the code further. 
Other options for further investigation include using different random distributions of input terminals, using structured input sets (to further test robustness), and finer time analysis to determine which part of the process takes more time, the OVD construction or the overlaying steps.

%
%The OOVD can be calculated in a reasonable amount of time on a modest computer for quite large point sets.
%For example, we were able to compute it for 10,000 random points in under 9 minutes on a 2.4GHz processor.
%For 1,000 points it took 30 seconds.
%

%Firtly, as we will discuss later, in practice we found that constructing the OOVD was much faster than the rest of the $1$-Steiner algorithm.

%In practice, even for the 1-Steiner case, it turns out that OOVD construction can be done very fast relative to the main part of the algorithm.

%This means that calculating the OOVD is easy, even when compared to later preprocessing steps, let alone the main part of the algorithm.
%Therefore there is little to be gained by using theoretically faster methods to calculate it, such as a sweep line algorithm.

%What can we say about typical complexity rather than worst case?

%Inspecting the examples we have calculated for uniformly distributed point sets, the number of edges (and thus faces) seems more like linear.
%So lets say the number of faces is $f$ for now.
%It seems plausible that the number of faces is linear in $n$, since one can imagine that only terminals within the local region of a terminal affect the number of faces related to that terminal. (Here I imagine the point set having a fixed density and larger sets occupy more space).
%Thus the number of faces per terminal may be essentially constant.

%%%%%%%%%%%%%%%%%%%%%
\section{$1$-Steiner trees as an application of OOVDs}\label{Steiner}
%%%%%%%%%%%%%%%%%%%%%

\begin{figure}
    \centering
    \includegraphics[width=\textwidth]{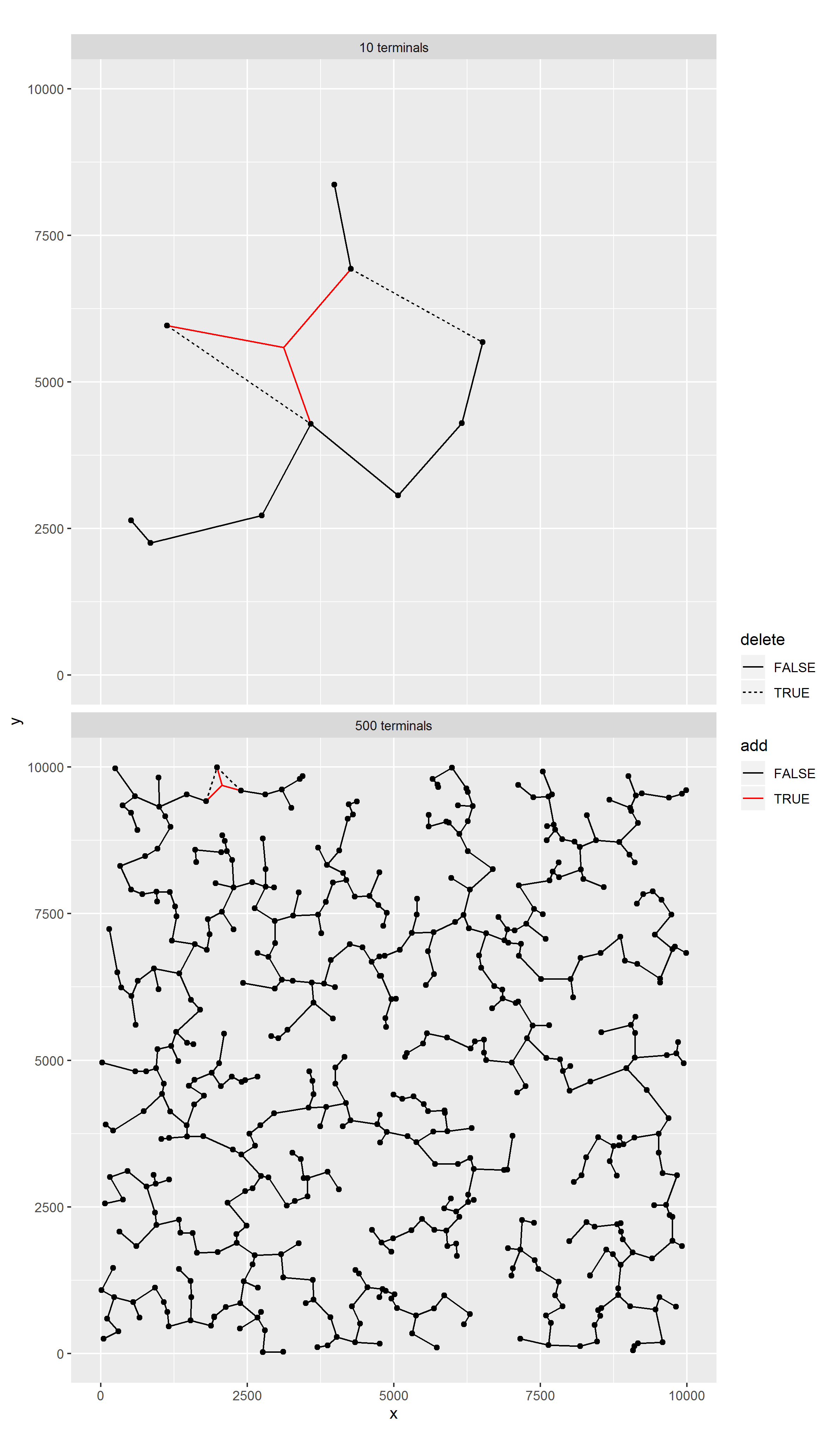}
    \caption{Minimum 1-Steiner trees for (top) 10 terminals, and (bottom) 500 terminals.}
    \label{fig:1steiner-examples}
\end{figure}

In this section we describe an algorithm, based on the one proposed in~\cite{GeorgeP87}, which employs the refined OOVD output in order to construct optimal Euclidean $1$-Steiner trees. We then analyse the improvement in efficiency that this provides over a more naive approach. % for $1$-Steiner tree construction.
Figure \ref{fig:1steiner-examples} shows two examples of optimal $1$-Steiner trees.

Constructing optimal $1$-Steiner trees is a polynomially solvable problem. This follows from two observations. Firstly, an optimal $1$-Steiner tree is an MST on its nodes. Therefore, if the optimal location of the Steiner point is known then an optimal $1$-Steiner tree can be constructed using Prim's algorithm. Secondly, if the neighbours of a Steiner point are known then the optimal location of the Steiner point, which coincides with the Fermat point of these neighbours, can be found in constant time. This is because a Steiner point never has degree more than $4$ in the Euclidean plane~\cite{RubinTW92}, %[this point was missed in papa] 
and the Fermat point of $3$ or of $4$ points can be exactly constructed in constant time~\cite{BrazilRT14}. %IS THIS THE RIGHT PAPER?

From the above observations, a naive algorithm for finding an optimal $1$-Steiner tree simply iterates through all sets of three or four terminals in $X$ and finds the Fermat point $s'$ of each set. An MST is then calculated on each set $X\cup \{s'\}$ and the shortest resulting tree is selected as the optimal solution. This naive approach has an asymptotic complexity of $O(n^5\log n)$, assuming that a Euclidean MST can be constructed in time $O(n\log n)$.

Employing the OOVD, the worst case complexity for $1$-Steiner tree construction can be reduced to $O(n^2)$ as follows: first the OOVD is created for the given set $X$ of input terminals and an MST $T'$ on $X$ is constructed. Next preprocessing is performed that stores the longest edge on every path in $T'$. This data will be used to update $T'$, in constant time, to include a Steiner point. For each region of the OOVD and its associated terminals, one selects a subset of at most $4$ terminals as neighbours for the Steiner point and then  calculates the optimal location of the Steiner point with respect to these neighbours. The MST is then updated in order to contain the Steiner point and its neighbours. Below we describe some of these steps in more detail and also present various refinements and practical improvements to this framework.    

\subsection{Calculating the Fermat point }
Suppose that $T$ is an an optimal $1$-Steiner tree on $X$ with Steiner point $s$, where $s$ has neighbour-set $N(s)=\{x_i: i\in I\}$ for some index set $I$. It is known that $s$ is of degree at most four~\cite{RubinTW92}, so %Clearly if $s$ is of degree less than $3$ then an MST on $X$ is an optimal $1$-Steiner tree. Therefore 
we may assume that $3\leq |I|\leq 4$.
Since $s$ minimises the sum of distances to its neighbours,
% expression $$F(s)=\displaystyle\sum_{i\in I}\|s-x_i\|.$$ In other words,
 $s$ is the \emph{Fermat point} of $N(s)$. Although there is no general algebraic solution for constructing Fermat points, there are simple constructions for three or four neighbours. 
% when $|I|\leq 4$ the point $s$ can be exactly constructed using simple geometric operations. 
If $|I|=3$ then $N(s)$ forms a triangle with all internal angles at most $120^\circ$ and where every pair of edges incident to $s$ meet at exactly $120^\circ$.
%For Steiner points of degree four we state the following well-known result . 
When $|I|=4$ the elements of $N(s)$ are in convex position and $s$ lies at the intersection of the two diagonals of $N(s)$~\cite{BrazilRT14}. %[cite marcus]

%However, a result of Rubenstein et al states that a FTST can't be minimal if it has degree five points, so we only need to consider degree three and four steiner points.
%Fortunately such points are easily constructible.

%Degree 3 are computed from the analytical solution presented in (Uteshev, 2014) - need to first check if any angle in the triangle is $\geq 120$ degrees.

%Degree 4 are at the intersection of two line segments. See Line-line intersection on Wikipedia.

\subsection{Processing the OOVD output}

Each face of the refined OOVD stores the closest terminal in each of the six cones plus the overall closest terminal.
This data is stored as a set $V$ of vectors that have 7 entries which are the indices of the relevant terminals in $X$ (or zero if there is no terminal in that cone), with the last entry for the closest terminal.
For a vector $v \in V$ we denote the associated face of the refined OOVD by $R(v)$.
%=(b_1,\ldots,b_7) \in B$
Thus if a point $s\in R(v)$ is a Steiner point in an optimal 1-Steiner tree, the indices of the neighbours of $s$ all occur in $v$.

The 1-Steiner algorithm involves iterating through every potential set of neighbours of the Steiner point that is to be added to $X$.
We call such a potential neighbour set a \emph{bucket}.
Naively, each subset of $3$ or $4$ of the first six entries of each $v\in V$ needs to be considered as a bucket, % in an optimal $1$-Steiner tree including $s$. 
yielding up to ${6 \choose 4} +{6 \choose 3}=35$ buckets per face.
In this subsection we describe a method of processing the OOVD data to produce a significantly smaller set of buckets $B$. 

%\begin{figure}
%    \centering
%    \includegraphics[width=8cm]{degree3.pdf}
%    \caption{}
%    \label{figdegree3}
%\end{figure}

%The processing consists of three main parts. 

Firstly, the local geometry of Steiner points allows us to ignore certain combinations of elements in each $v\in V$ which cannot correspond to the neighbours of a Steiner point. 
Take any $v=(v_1,\ldots,v_7) \in V$ %, let $R(v)$ be the region of the refined OOVD associated with $b$, 
and suppose $s$ is an optimal Steiner point lying in $R(v)$. 
If $s$ is of degree $3$,
% in an optimal $1$-Steiner tree $T$ and let $x_{v_i},x_{v_j},x_{v_k}$ be the neighbours of $s$ in $T$, where $1\leq i<j<k\leq 6$. 
since each cone $C_i$ has an angle of $60^\circ$ and the edges incident to a degree-$3$ Steiner point meet at $120^\circ$, %we must have $k=j+2$ and $j=i+2$; see Figure \ref{figdegree3}. Therefore $b$ only contains 
there are only two feasible subsets of size $3$, % from its first six components, 
namely $\{v_1,v_3,v_5\}$ and $\{v_2,v_4,v_6\}$.
If $s$ is of degree $4$ then it lies at the intersection of the two diagonals of the quadrilateral formed by its neighbours.
Therefore the neighbours of $s$ are two pairs of terminals that occur within opposite cones %In other words, these pairs must come from the set $\{(b_1,b_4), (b_2,b_5), (b_3,b_6)\}$.
and so the feasible subsets of size $4$ in $v$ are $\{v_1,v_2,v_4,v_5\}$, $\{v_1,v_3,v_4,v_6\}$ and $\{v_2,v_3,v_5,v_6\}$.
Each face therefore yields at most $5$ buckets in $B$. 

%The second processing step employs information from the classical Voronoi diagram on $X$. %This information is stored in the seventh component of each bucket $b\in B$. 

Moreover, since the Steiner point in an optimal 1-Steiner tree must be adjacent to its closest neighbour in $X$, % (because it is an MST on $X \cup \{s\}$).
%
%the following observation.
%
%\begin{lemma}Let $T$ be an optimal $1$-Steiner tree on $X$ and let $s$ be the Steiner point of $T$. Let $x\in X$ be any terminal so that $\|s-x\|\leq \|s-y\|$ for all $y\in X$. Then there exists an optimal $1$-Steiner tree $T'$ on $X$ that contains edge $xs$. 
%\end{lemma}
%\begin{proof}
%Suppose that the neighbour-set of $s$ in $T$ is $N(s)=\{x_i: i\in I\}$ for some index set $I$, and let $x\in X$ any terminal so that $\|s-x\|\leq \|s-y\|$ for all $y\in X$. If $x\in N(s)$ then the result follows. Therefore suppose $x\notin N(s)$. Adding edge $xs$ to $T$ produces a cycle containing one of the $x_i$, say (without loss of generality) $x_1$. Deleting edge $x_1s$ from this cycle produces a tree spanning $X\cup \{s\}$ that is no longer than $T$, and is therefore optimal.
%\end{proof}
%
%
%Now consider some $b\in B$ where $b=(b_1,...,b_7)$. 
%If $s\in R(b)$ of the refined OOVD diagram then $x_{b_7}$ is a closest terminal to $s$. 
%Therefore, any feasible subset of indices of terminals adjacent to $s$ in an optimal solution can be assumed to contain $b_7$. 
%
%This means 
any bucket derived from $v$ can be eliminated if it does not contain $v_7$.
This eliminates one of the 3-entry buckets, and one of the 4-entry buckets, leaving at most $3$ buckets per face.

Finally, note that it is possible for the same bucket to arise from multiple faces of the refined OOVD. The final stage of processing consists of removing such repetitions.

%overlaying the classical Voronoi diagram with the OOVD subdivides the faces of the OOVD. Therefore, some buckets may appear more than once in the set $B^*$. The third stage of processing consist of removing any such repetitions.

\subsection{Preprocessing for the MST update step}
One of the steps in the algorithm for constructing an optimal $1$-Steiner tree consists of updating a MST to include a new point $s$. This step can be done in constant time if suitable preprocessing has been performed. This preprocessing step calculates a minimum spanning tree $T'$ on $X$, and then for each pair of terminals stores in a table $H$ the longest edge on the path between these nodes in the MST. The preprocessing step can be perfomed in a time of $O(n^2)$ [ref GenkStei].

\subsection{The $1$-Steiner algorithm}

We now present pseudo-code for the $1$-Steiner algorithm. 

\begin{algorithm}\label{algFPT}
\caption{(The $1$-Steiner tree algorithm)}
\begin{algorithmic}[1] 
\REQUIRE A set of terminals $X$ in the plane and the refined OOVD. 
\ENSURE A point $s\in \mathbb{R}^2$ and a tree $T$ spanning $X\cup\{s\}$ such that  $\sum_{e\in E(T)} \|e\|$ is minimised.
%\bigskip
\STATE {Process the refined OOVD data $V$ to get the reduced bucket list $B$.}
\STATE {Construct an MST $T'$ on $X$.}
\STATE {For every pair of nodes $u,v$ of $T'$, find the longest edge on the path connecting $u$ and $v$ in $T'$. Store the results in a table $H$. }
\FOR{each bucket $b\in B$}
\STATE {Construct the Fermat point $s(b)$.}
\STATE Add the edge $x_is(b)$ to $T'$ for each $x_i\in b$. Then, employing $H$, delete the longest edge on the path connecting $x_i$ and $x_j$ in $T'$ for each \{$x_i,x_j\}\subset b$. Let $T(b)$ be the resultant tree.
\ENDFOR
\STATE{Let $T$ be the shortest $T(b)$ found.}
\end{algorithmic}
\end{algorithm}

The complexity of the $1$-Steiner tree algorithm is $O(n^2)$. This follows from the fact that Line 1 can be performed in $O(n^2)$ time, since $B$ contains $O(n^2)$ buckets. Then, Line 2 requires $O(n\log n)$ time and Line $3$ requires $O(n^2)$ time. The For loop in Line 4 runs at most $O(n^2)$ times, and in each execution of the loop the Fermat point is constructed and the MST updated in constant time.

\subsection{Experimental results for $1$-Steiner trees}\label{sub1SteinerExp}
In this section we provide experimental results on the effect that the OOVD face data preprocessing has on the number of buckets that must be considered during $1$-Steiner tree construction. 
Recall that construction time is directly proportional to the number of buckets considered since updating an MST and constructing a Fermat point are done in constant time.
We also provide some statistical observations on properties of the optimal $1$-Steiner trees that arose.

%Since construction time is directly proportional to the number of neighbour-sets considered (since updating an MST and constructing a Fermat point are done in constant time), we will only look at [ratios of the number of buckets] [or say something like: we will not be performing experiments on direct construction of $1$-Steiner trees. Rather, we will analyse the effect in terms of the reduction in buckets.].

\begin{figure}
    \centering
    \includegraphics[width=\textwidth]{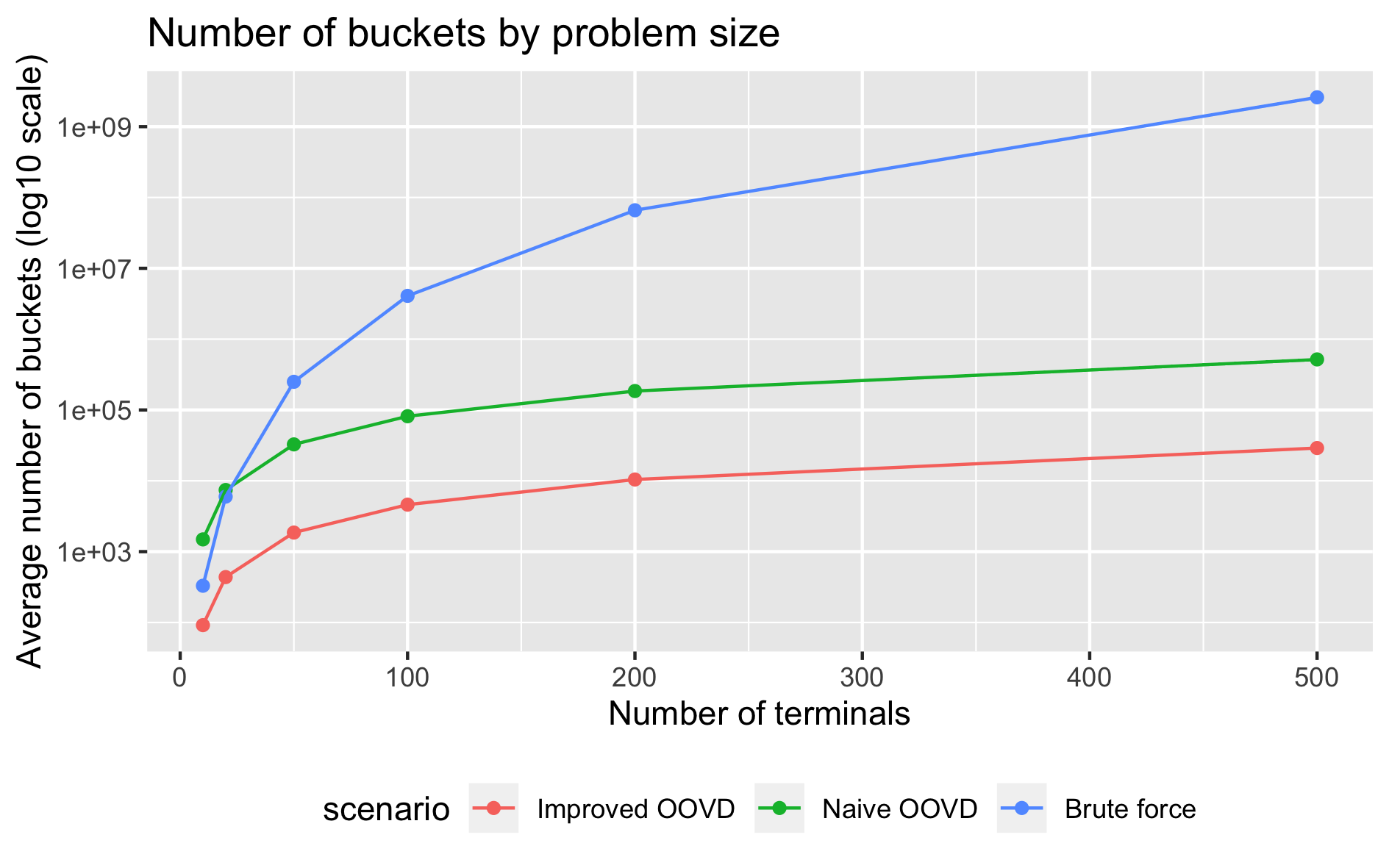}
    \caption{Number of buckets by problem size (averages of 50 seeds).}
    \label{fig:buckets}
\end{figure}

Figure \ref{fig:buckets} shows the number of buckets generated for three scenarios. The blue curve represents %the number of buckets considered %for the instances generated in Section [oovd experiments] 
the brute force approach, with no use of OOVD data. In other words, this is simply a plot of the function ${n \choose 4}+{n\choose 3}$.
The green curve represents the number of buckets considered when employing OOVD data in a naive way, i.e., with 35 buckets per face. %In other words, in the naive OOVD method every set of $3$ or $4$ terminals is considered as a potential neighbour set for each bucket in $B$. 
Finally, the orange curve represents the number of buckets in $B$ which resulted from processing the refined OOVD data $V$ as discussed above.

It can be seen that employing OOVDs results in a significant reduction in the number of buckets that need to be condidered. In fact, for $500$ terminals this reduction is of the order $10^5$ over the brute force approach.
It also shows that the improved OOVD approach provides at least a factor 12 improvement over the simple OOVD scenario. This is as expected since the reduction from 35 to 3 buckets per face provides nearly a factor 12 in itself.

%In Figure we compare the naive OOVD method to our improved OOVD processing more closely. Here it can be seen that...

%\begin{figure}
%    \centering
%    \includegraphics[width=\textwidth]{fig-buckets-2.png}
%    \caption{Number of buckets by problem size (averages of 50 seeds).}
%    \label{fig:buckets2}
%\end{figure}

%\subsection{Statistical and geometric properties of minimum 1-Steiner trees}

We now mention a number of statistical properties of optimal $1$-Steiner trees that we observed for trees generated in these experiments.

\begin{figure}
    \centering
    \includegraphics[width=\textwidth]{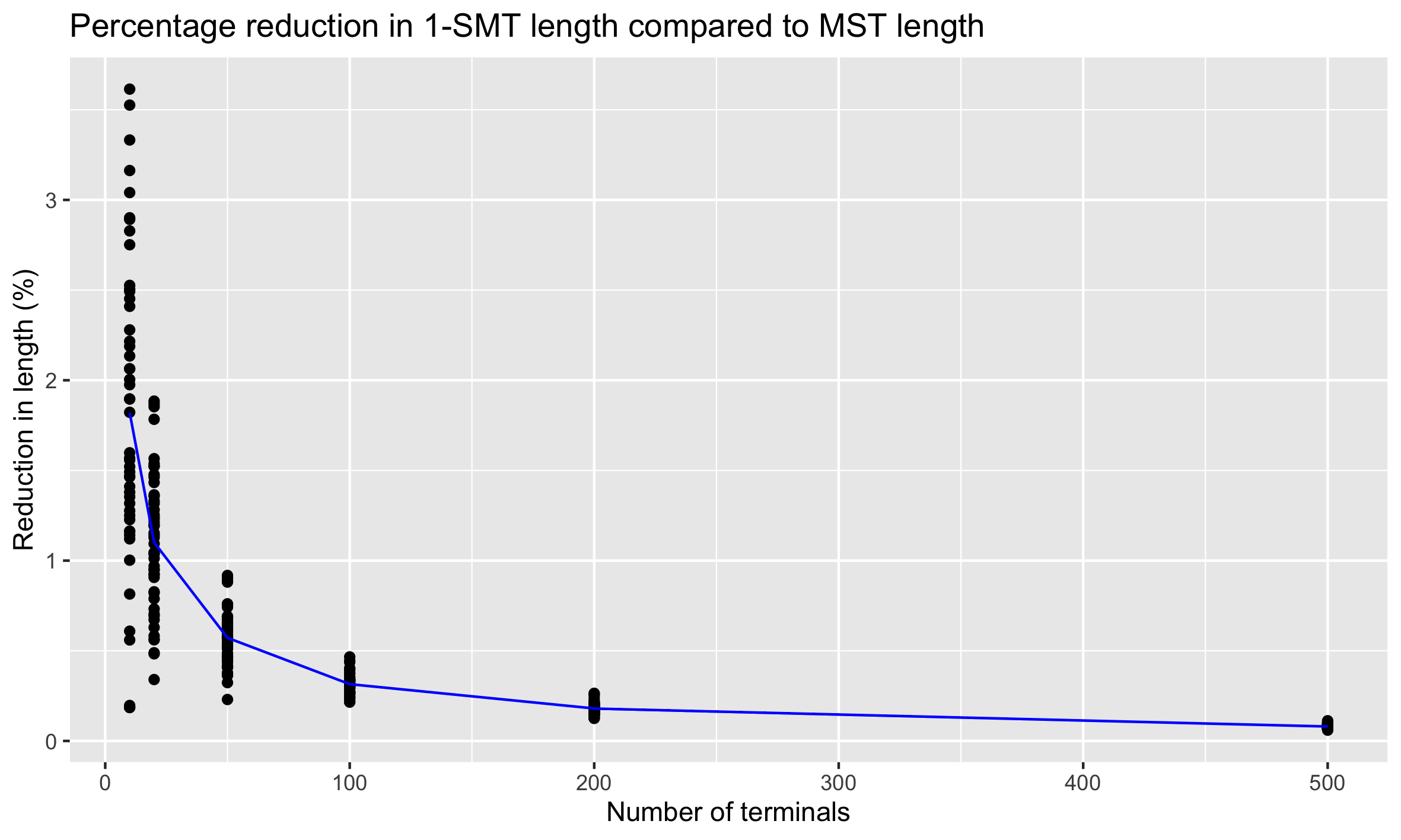}
    \caption{Percentage reduction in 1-SMT length compared to MST length.}
    \label{fig:improvement}
\end{figure}

\begin{figure}
    \centering
    \includegraphics[width=\textwidth]{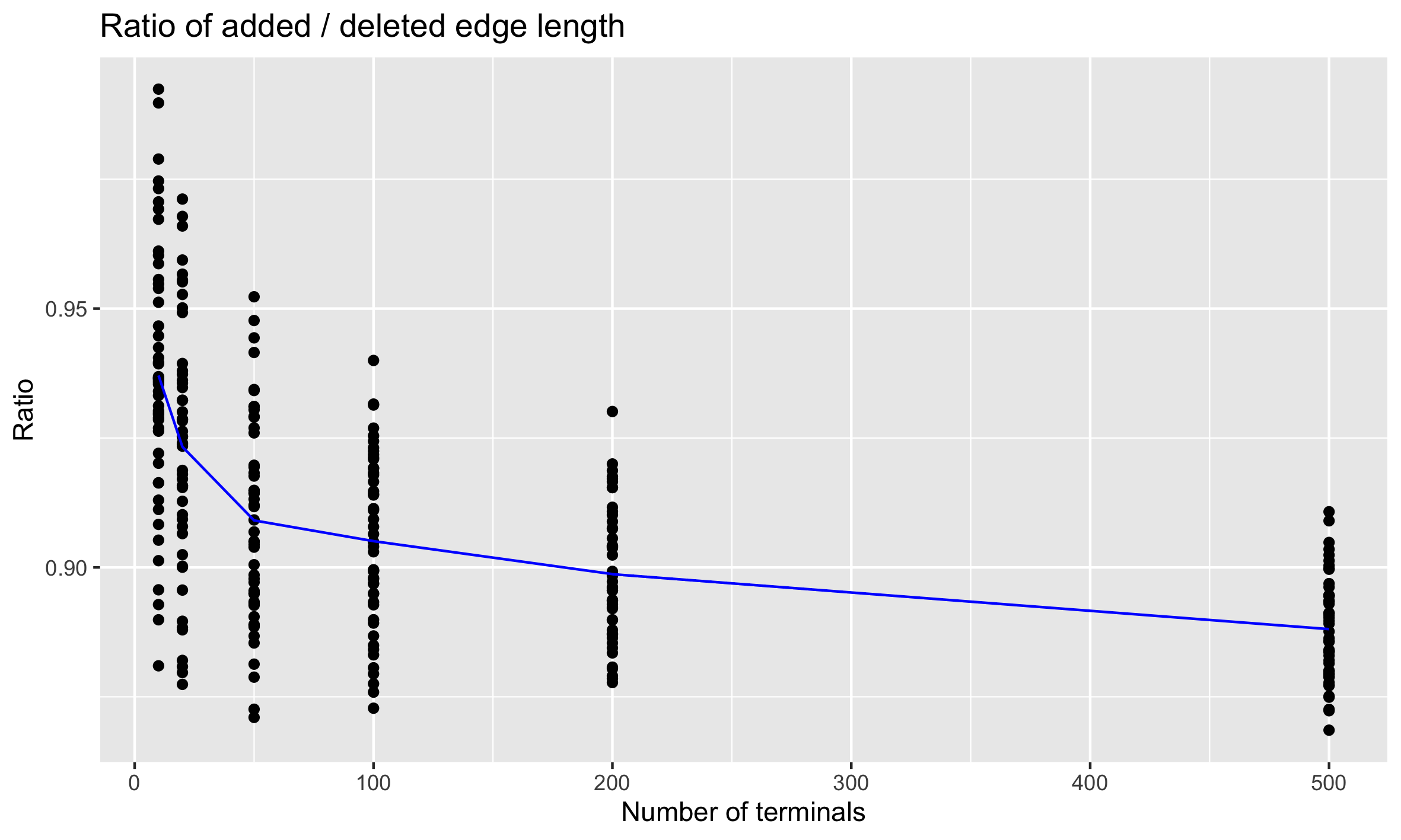}
    \caption{The ratio of added / deleted edge length.}
    \label{fig:add_delete_ratio}
\end{figure}

Firstly, in Figure \ref{fig:improvement} we have plotted the percentage reduction in network length that is achieved by employing a single Steiner point versus the MST. As is expected, this percentage decreases as $n$ increases. In Figure \ref{fig:add_delete_ratio} we present a local measure of improvement in length by comparing the length of the Steiner edges added to the MST and the MST edges deleted.
       
\begin{figure}
    \centering
    \includegraphics[width=\textwidth]{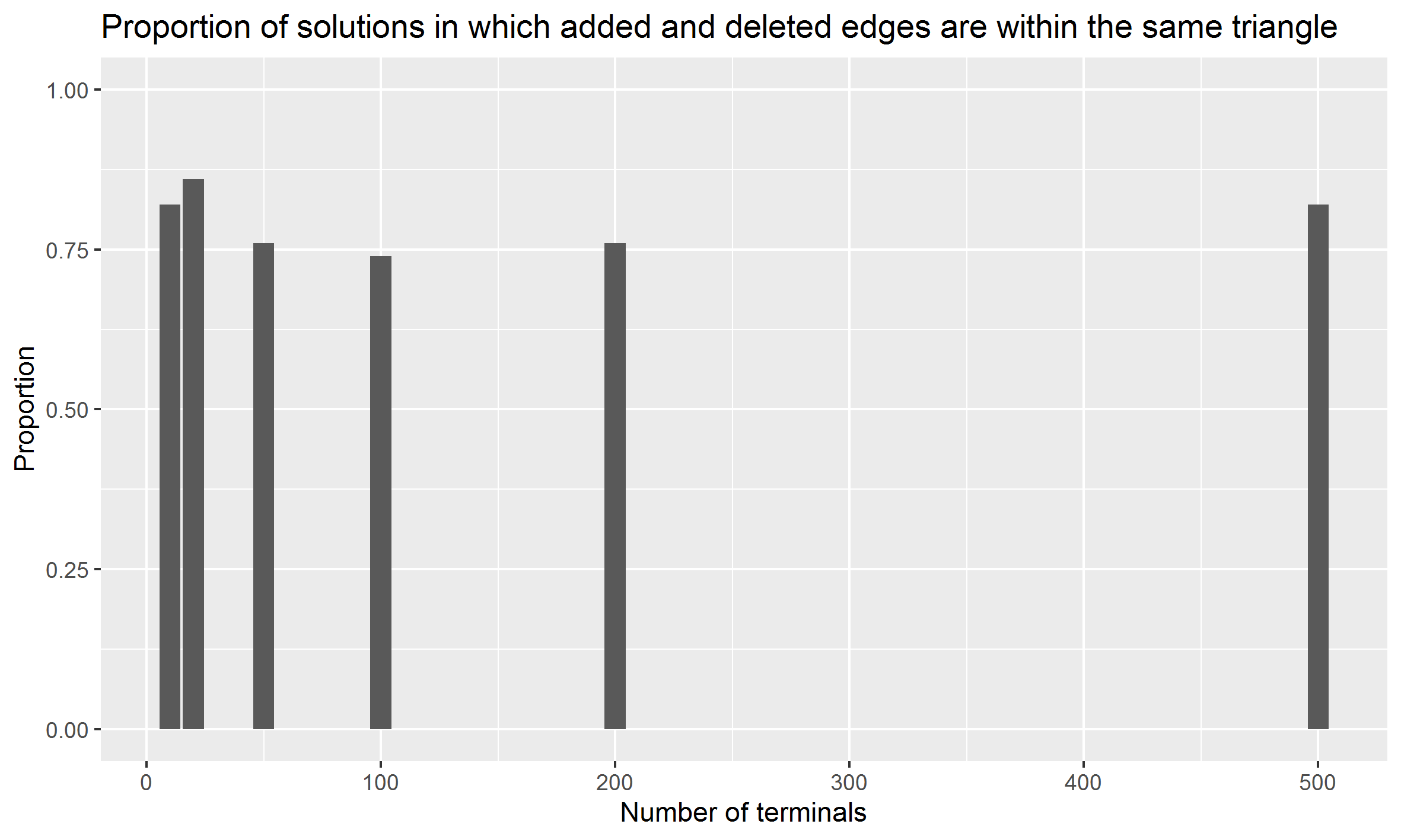}
    \caption{The proportion of solutions in which the added and deleted edges are from the same triangle.}
    \label{fig:prop_triangle}
\end{figure}

We also observed that in many cases the optimal $1$-Steiner tree results from deleting a pair of adjacent edges in the MST. In Figure \ref{fig:prop_triangle} we have plotted the proportion of times this occurred for our instances. As can be seen, our experiments suggest that this occurs on average between $75\%$ and $88\%$ of the time for uniformly generated instances.

\begin{figure}
    \centering
    \includegraphics[width=\textwidth]{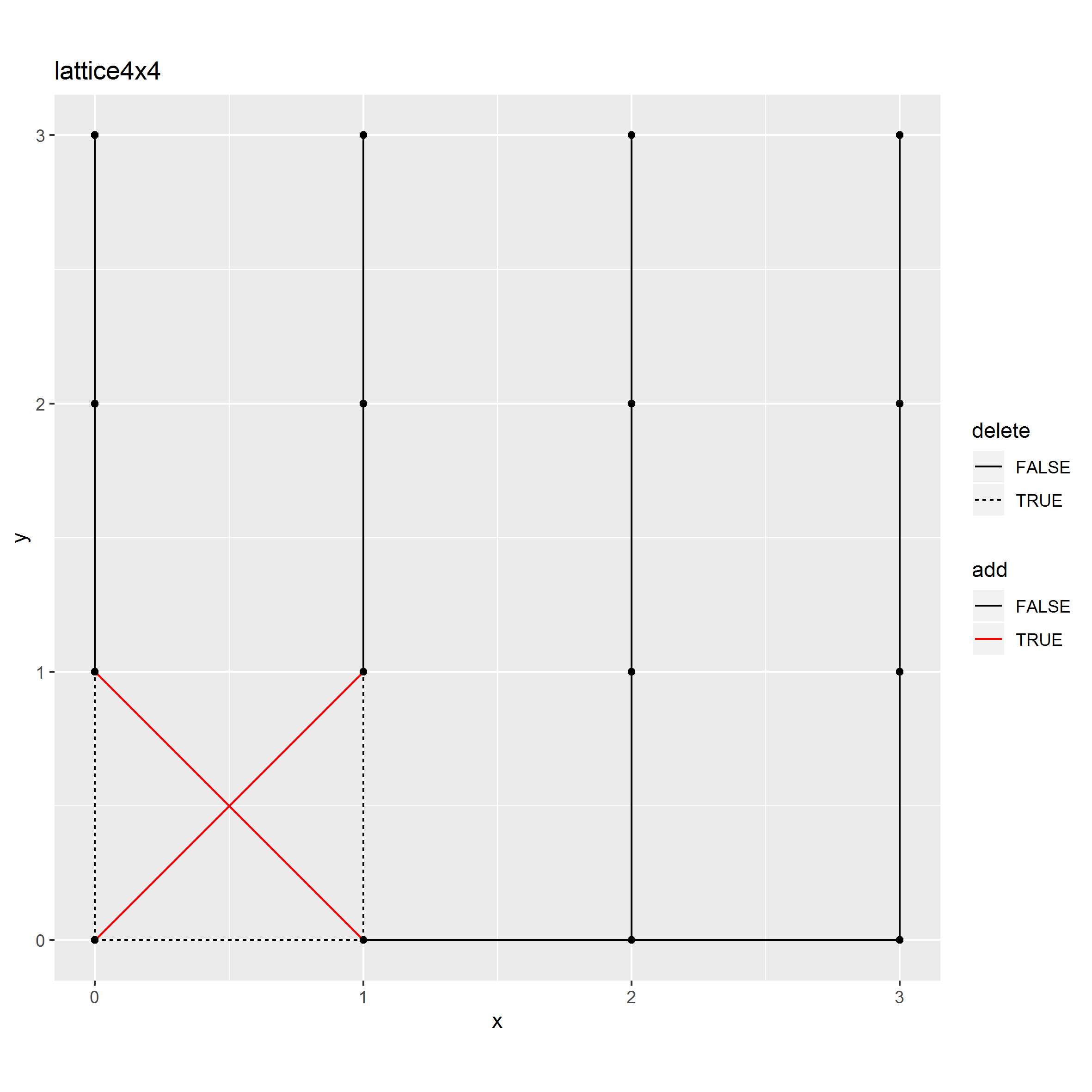}
    \caption{The solution for a lattice problem instance contains a cross.}
    \label{fig:1steiner_lattice4x4}
\end{figure}

Finally, even though degree-$4$ Steiner points are possible in an optimal $1$-Steiner trees, our experiments show that they are rare. In fact, not a single instance we constructed from the random examples contained a degree-$4$ Steiner point. Figure \ref{fig:1steiner_lattice4x4} shows an example of an optimal tree on a contrived instance which does contain a degree-$4$ point.

\bibliography{bib}
\bibliographystyle{plain}
\end{document}